\documentclass[a4paper,useAMS,usenatbib]{mn2e}

\usepackage{dcolumn,graphicx,multirow}
\usepackage[varg]{txfonts}
\usepackage[T1]{fontenc}
\usepackage{aecompl}

\newcommand{\iras}{IRAS16293}
\newcommand{\kms} {\ensuremath{\rm km~s^{-1}}}
\newcommand{\vlsr}{\ensuremath{V_{\mathrm{LSR}}}}

\newcommand{\cc}  {cm$^{-3}$}
\newcommand{\cs}  {cm$^{-2}$}
\newcommand{\tex}{\ensuremath{T_{\rm ex}}}
\newcommand{\tcmb}{\ensuremath{T_{\rm CMB}}}
\newcommand{\e}{\ensuremath{{\rm e}}}

\newcommand{\phn}{\phantom{0}}

\newenvironment{myitemize}
	{\begin{itemize}
		\setlength{\leftmargin}{2.75cm}}
	{\end{itemize}}
      
\graphicspath{{./}
{./figures/}
}

\title[CH in absorption in IRAS16293$-$2422]{CH in absorption in IRAS16293$-$2422 \thanks{based on observations with {\it Herschel}/HIFI. {\it Herschel} is an ESA space observatory with science instruments provided by European-led Principal Investigator consortia and with important participation from NASA.}}
\author[S. Bottinelli et al.]{S. Bottinelli,$^{1,2}$\thanks{E-mail:sbottinelli@irap.omp.eu}
V. Wakelam,$^{3,4}$
E. Caux,$^{1,2}$
C. Vastel,$^{1,2}$
Y. Aikawa$^{5}$
and C. Ceccarelli$^{6}$
\\
$^{1}$ Universit\'e de Toulouse; UPS-OMP; Institut de Recherche en Astrophysique et Plan\'etologie (IRAP) ; UMR 5277 ; Toulouse, France\\
$^{2}$ CNRS; IRAP; UMR 5277 ; 9 Av. colonel Roche, BP 44346, F-31028 Toulouse cedex 4, France\\
$^{3}$ Univ. Bordeaux, LAB, UMR 5804, F-33270 Floirac, France \\
$^{4}$ CNRS, LAB, UMR 5804, F-33270 Floirac, France \\
$^{5}$ Department of Earth and Planetary Sciences, Kobe University, Kobe 657-8501, Japan \\
$^{6}$ UJF-Grenoble 1 / CNRS-INSU, Institut de Plan\'etologie et d'Astrophysique de
Grenoble (IPAG) UMR 5274, Grenoble F-38041, France \\
}
\begin{document}

\date{}

\pagerange{\pageref{firstpage}--\pageref{lastpage}} \pubyear{2013}

\maketitle

\label{firstpage}

\begin{abstract}
While recent studies of the solar-mass protostar IRAS16293$-$2422 have focused on its inner arcsecond, 
the wealth of {\it Herschel}/HIFI data has shown that the structure of the outer envelope and of the transition region to the more diffuse ISM is not clearly constrained. 
We use rotational ground-state transitions of CH (methylidyne),
as a tracer of the lower-density envelope.
Assuming LTE, we perform a $\chi^2$ minimization of the high spectral resolution HIFI observations of the CH transitions at $\sim$532 and $\sim$536~GHz in order to derive column densities in the envelope and in the foreground cloud.
We obtain column densities of $(7.7\pm0.2)\times10^{13}$~\cs\ and $(1.5\pm0.3)\times10^{13}$~\cs, respectively. The chemical modeling predicts column densities of $(0.5-2)\times10^{13}$~\cs\ in the envelope (depending on the cosmic-ray ionization rate), and $5\times10^{11}$ to $2.5\times10^{14}$~\cs\ in the foreground cloud (depending on time).\\
Both observed abundances are reproduced by the model at a satisfactory level.  The constraints set by these observations on the physical conditions in the foreground cloud are however weak.
Furthermore, the CH abundance in the envelope is strongly affected by the rate coefficient of the reaction H+CH$\rightarrow$C+H$_2$ ; 
further investigation of its value at low temperature would be necessary to 
facilitate the comparison between the model and the observations.
\end{abstract}


\section{Introduction}
\label{sec: intro}

IRAS16293$-$2422 (hereafter \iras) is a well-known close-by ($d$ = 120 pc), low-mass ($\sim1M_\odot$), Class 0 protostar
which has been the target of many studies, revealing its complex structure and composition:
binarity and even multiplicity of the source \citep[e.g.][]{mundy-etal86,wootten89}, 
presence of multiple shocks and outflows \citep[e.g.][]{mizuno-etal90,walker-etal90,castets-etal01,stark-etal04,yeh-etal08},
rich spectra, in particular containing complex organic molecules \citep{blake-etal94,vandishoeck-etal95,cazaux-etal03,caux-etal11},
a fact that earned it the title of first hot corino, in analogy to its massive counterparts, hot cores.

Over the past few years, studies of \iras\ have mostly focused on 
the chemical richness and complexity of the hot corino and on imaging the inner arcseconds of the source 
\citep[e.g., ][]{jorgensen-etal11,bisschop-etal08-i16293,takakuwa-etal07,chandler-etal05,bottinelli-etal04-iras16293,kuan-etal04}.
However, it has recently appeared that the structure of the outer envelope is also not very well known, in particular
the transition region from the envelope to the more diffuse interstellar medium (ISM). 
For example, \citet{coutens-etal12} find that the presence of 
an additional layer of gas in the line of sight of the protostar is necessary to explain the self-absorption observed in two HDO fundamental lines. They called this additional component  a ``photo-evaporation'' or ``photo-desorption'' layer, assuming that photo-evaporation processes, induced by the UV radiation field corresponding to a visual extinction between 1 and 4, would be responsible for the larger abundances of HDO observed in this component \citep{hollenbach-etal09}. We will give it a more general name, ``foreground cloud'', to avoid any predefined idea on its chemistry.

In this paper, we take advantage of the unique capability of the Heterodyne Instrument for the Far-Infrared (HIFI, \citealt{degraauw-etal10})
onboard the {\it Herschel} Space Observatory \citep{pilbratt-etal10} to observe the rotational ground-state transitions of a light diatomic hydride,
CH (methylidyne). The importance of CH 
resides mostly in the fact that it is a key species of the chemical networks: CH is the first and simplest neutral molecule formed from C and/or C$^+$ in the gas phase and an important chemical intermediate in the production pathway of CO from C and C$^+$ ; it is also suggested to be a key reactant in the production of carbon-chain molecules at early evolutionary stages \citep[e.g., ][]{sakai-etal07}.
 \\
In section \ref{sec: obs}, we present observations of the ground-state transitions of CH obtained with HIFI. We analyze these data with an LTE model in section \ref{sec: lte} and a chemical modeling is performed in section \ref{sec: chem}.

\section{Observations and data reduction}
\label{sec: obs}

\begin{table*}
\begin{minipage}{155mm}
\caption{Observational and spectroscopic parameters. \label{tab: obs}}
\begin{tabular}{ccccccc cc}
\hline\hline
HIFI  & Transition  & Frequency & A$_{u\ell}$ & E$_{\rm up}$ & rms$^a$ & Continuum$^a$ & HPBW & Beam   \\
band& ($N$, $J$, parity, $F$ $\leftarrow N'$, $J'$, parity , $F'$) & (GHz) & (s$^{-1}$) & (K)  & (mK) & (K) &$('')$ & efficiency \\
\hline
1a 		& CH (1 3/2 $-$ 1 $\leftarrow$ 1 1/2 $+$ 1) & 532.7217 & 2.1$\times10^{-4}$ & \multirow{3}{*}{25.7} & \multirow{3}{*}{11 (16)} & \multirow{6}{*}{0.22} &\multirow{6}{*}{40} & \multirow{6}{*}{0.76} \\
1a 		& CH (1 3/2 $-$ 2 $\leftarrow$ 1 1/2 $+$ 1) & 532.7239 & 6.2$\times10^{-4}$ & &  & & & \\
1a 		& CH (1 3/2 $-$ 1 $\leftarrow$ 1 1/2 $+$ 0) & 532.7933 & 4.1$\times10^{-4}$ & &  & & & \\
1a 		& CH (1 3/2 $+$ 2 $\leftarrow$ 1 1/2 $-$ 1) & 536.7611 & 6.4$\times10^{-4}$ & \multirow{3}{*}{25.8} & \multirow{3}{*}{11 (14)} & & & \\
1a 		& CH (1 3/2 $+$ 1 $\leftarrow$ 1 1/2 $-$ 1) & 536.7820 & 2.1$\times10^{-4}$ & &  & & & \\
1a 		& CH (1 3/2 $+$ 1 $\leftarrow$ 1 1/2 $-$ 0) & 536.7957 & 4.3$\times10^{-4}$ & &  & & & \\
\hline
\multicolumn{9}{c}{Non-detections$^b$}  \\
\hline
6b & CH (2 5/2 $-$ * $\leftarrow$ 1 3/2 $+$ *) & $\sim$ 1657.0   & 3.7$\times10^{-2}$ & 105.3 & 207 & \multirow{2}{*}{1.79} & \multirow{2}{*}{13} & \multirow{2}{*}{0.72}  \\
6b & CH (2 5/2 $+$ * $\leftarrow$ 1 3/2 $-$ *) & $\sim$ 1661.1   & 3.8$\times10^{-2}$ & 105.5 &  235 &  & &  \\
3b & CD (2 3/2 $-$ * $\leftarrow$ 1 3/2 $+$ *) & \phn$\sim$ 884.8 & 7.9$\times10^{-4}$ & 65.5 & 28 & \multirow{2}{*}{0.75} & \multirow{4}{*}{24} & \multirow{4}{*}{0.75}  \\
3b & CD (2 3/2 $+$ * $\leftarrow$ 1 3/2 $-$ *) & \phn$\sim$ 887.2 & 7.9$\times10^{-4}$ & 65.6 & 38 &  & & \\
3b & CD (2 5/2 $-$ * $\leftarrow$ 1 3/2 $+$ *) & \phn$\sim$ 915.9 & 5.6$\times10^{-3}$ & 67.0 & 37 & \multirow{2}{*}{0.80} & & \\
3b & CD (2 5/2 $+$ * $\leftarrow$ 1 3/2 $-$ *) & \phn$\sim$ 917.0 & 6.7$\times10^{-3}$ & 67.1 & 42 &  & &  \\
1a & $^{13}$CH (1 3/2 $+$ * $\leftarrow$ 1 1/2 $-$ *) & \phn$\sim$ 532.1 & 4.8$\times10^{-4}$ & 25.7 & 9 & \multirow{2}{*}{0.22} &\multirow{2}{*}{40} & \multirow{2}{*}{0.76} \\
1a & $^{13}$CH (1 3/2 $-$ * $\leftarrow$ 1 1/2 $+$ *) & \phn$\sim$ 536.0 & 4.7$\times10^{-4}$ & 25.7 & 10 & & & \\
6b & $^{13}$CH (2 5/2 $-$ * $\leftarrow$ 1 3/2 $+$ *) & \phn$\sim$ 1647.0 & 3.4$\times10^{-2}$ & 104.7 & 265 & \multirow{2}{*}{1.79} & \multirow{2}{*}{13} & \multirow{2}{*}{0.72} \\
6b & $^{13}$CH (2 5/2 $+$ * $\leftarrow$ 1 3/2 $-$ *) & \phn$\sim$ 1651.2 & 3.4$\times10^{-2}$ & 105.0 & 272 & & & \\
\hline
\end{tabular}

 \medskip
$^a$ On main-beam temperature scale ; rms is given in 0.5 MHz channels (WBS data) ; 
the rms of the pointed HRS observations is indicated in parentheses and is given in 240 and 60 kHz channels, at $\sim$532.7 and $\sim$536.8~GHz respectively.\\ 
$^b$ For non-detections, we only indicate the approximate frequency of all hyperfine transitions (different values of $F$, in this case represented by a star, *)
and give the highest A$_{u\ell}$.\\
{\sc Notes.}--- (i) The CD($N$=1--1) hyperfine transitions are around 439 GHz, not observable by HIFI.
(ii) Other hyperfine transitions of CH(2--1) exist around 1470.7 and 1477.3 GHz (band 6a), but were not observed
since only a partial survey of band 6a was carried out toward \iras\ and it did not cover these frequencies.

\end{minipage}
\end{table*}

\subsection{Survey data}

\begin{figure}
\includegraphics[angle=90,width=\columnwidth]{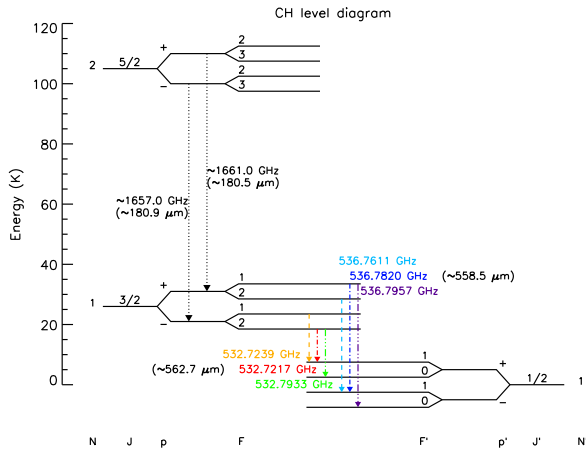}
\caption{Energy-level diagram of the lower rotational levels of 
of CH, 
with the $\Lambda$-doubling and hyperfine structure exaggerated 
for clarity; total parity is given by the plus and minus signs. The hyperfine-split $\Lambda$-doublet transitions studied in this article are shown as colored arrows : for each triplet, transitions with the lowest, intermediate and highest frequencies are represented by dashed, dot-dashed and dot-dot-dashed arrows. The non-detected, higher-frequency transitions are indicated as dotted arrows. For all groups of transitions, the corresponding approximate wavelength is given in parentheses.
\label{fig: nrj levels}}
\end{figure}

The solar type mass protostar \iras\ was observed 
as part of the HIFI guaranteed time Key Program CHESS \citep{ceccarelli-etal10}. 
The data presented in this article are part of a full spectral coverage of bands 
1a (480 -- 560\,GHz ; Obs. Id 1342191499) and 3b (858 -- 961\,GHz ; Obs. Id 1342192330), and a partial coverage of band 6b (1573 -- 1703\,GHz ; Obs. Id 1342191794) 
which were performed on March 1$^{\rm st}$, 19$^{\rm th}$, and 8$^{\rm th}$ 2010, respectively, 
using the HIFI Spectral Scan Double Beam Switch (DBS) observing mode with optimization of the continuum. 
In this mode, the HIFI acousto-optic Wide Band Spectrometer (WBS) was used, providing a spectral resolution of 1.1~MHz 
($\sim$0.6~\kms\ at 520~GHz and $\sim$0.3~\kms\ at 1~THz) 
over an instantaneous bandwidth of 4$\times$1\,GHz \citep{roelfsema-etal12}. 
The observed coordinates were $\alpha_{2000}$ = 16$^{\rm h}$ 32$^{\rm m}$ 22$\fs$64, $\delta_{2000}$ = $-$ 24$\degr$ 28$\arcmin$ 33$\farcs$6. 
The DBS reference positions were situated approximately $3\arcmin$ east and west of the source. 
Table~\ref{tab: obs} summarizes the observation parameters as well as spectroscopic information from \citet{amano00} and from the JPL \citep{pickett-etal98} and CDMS \citep{muller-etal01,muller-etal05} databases. The latter made use of data from \citet{davidson-etal01}, \citet{mccarthy-etal06}, and \citet{phelps+dalby66} for CH, \citet{halfen-etal08} for CD, and \citet{halfen-etal08} and \citet{mccarthy-etal06} for $^{13}$CH.
Energy levels are shown in Fig.~\ref{fig: nrj levels}.

The data have been processed using the standard HIFI pipeline up to frequency and intensity calibrations (level 2) with the ESA-supported package HIPE 10 
\citep{ott-etal10}. 
A single local-oscillator-tuning spectrum consists, for each polarization, in 4 sub-bands of $\sim\,1$~GHz for the SIS bands (1 to 5), and in 1 sub-band of $\sim\,0.4$~GHz and 2 sub-bands of $\sim~1$~GHz for the HEB bands (6 and 7).\\
Using a routine developed within the HIFI ICC (Instrument Control Center), {\it flagTool}, 
we examined each sub-band in order to remove spurs that were not automatically removed by the pipeline.
Removal of standing waves was performed with the HIPE task {\it fitHifiFringe}
and {\it fitBaseline} was used to fit a low-order polynomial baseline to line-free channels. 
Sideband deconvolution was performed with the dedicated HIPE task {\it doDeconvolution}.

The spectra observed in both horizontal and vertical polarization were of similar quality, and averaged to lower the noise in the final spectrum. 
This is justified since polarization is not a concern for the presented analysis. 
The continuum values obtained from running {\it fitBaseline} are well fitted by polynomials of order 3 
over the frequency range of the whole band. 
The single side band continuum derived from the polynomial fit at the considered frequency (Table~\ref{tab: obs}) was eventually added to the spectra. 
For the analysis, intensities are then converted from antenna to main-beam temperature scale using a forward efficiency of 0.96 and
the (frequency-dependent) beam-efficiency taken from \citet{roelfsema-etal12} and reported in Table~\ref{tab: obs}. 
The final spectra are displayed in the top panels of Fig.~\ref{fig: ch hfs fit}.

\subsection{Pointed observations}

A dedicated pointed observation of the CH transition around 536~GHz was performed on February 15$^{\rm th}$ 2011 (Obs. Id 1342214339), 
for which the High-Resolution Spectrometer (HRS) was used in addition to the WBS, 
providing the best possible spectral resolution for this instrument, namely, 120~kHz ($\sim$0.07~\kms\ at 536~GHz). 
The on-source integration time was 26 minutes, and the LO frequency was 531.980~GHz. 

Additionally, we search the {\it Herschel} Science Archive (HSA) and found two pointed datasets (Obs. Id 1342227403 and 1342227404) covering the $\sim$532-GHz transition,
acquired on August 26$^{\rm th}$ 2011, but for which the HRS resolution was only 480~kHz ($\sim$0.27~\kms\ at 532~GHz). 
The on-source integration times were 2.5 and 2.8 minutes, and the LO frequencies were 526.266 and 538.3665~GHz, respectively. 

In both cases, 
the observed coordinates were less than an arcsecond away from that of the survey data.
All pointed data were also processed using the standard HIFI pipeline up to frequency and intensity calibrations (level 2) with HIPE 10.
For each transitions, the integrations were weighted-averaged based on their individual rms determined by fitting a low-order polynomial baseline to line-free channels.
As for the survey data, both polarizations were averaged and intensities were converted to main-beam temperature scale.
The final spectra are displayed in the bottom panels of Fig.~\ref{fig: ch hfs fit}.

\section{Results and analysis}
\label{sec: lte}

\begin{figure*}
\includegraphics[width=0.95\columnwidth]{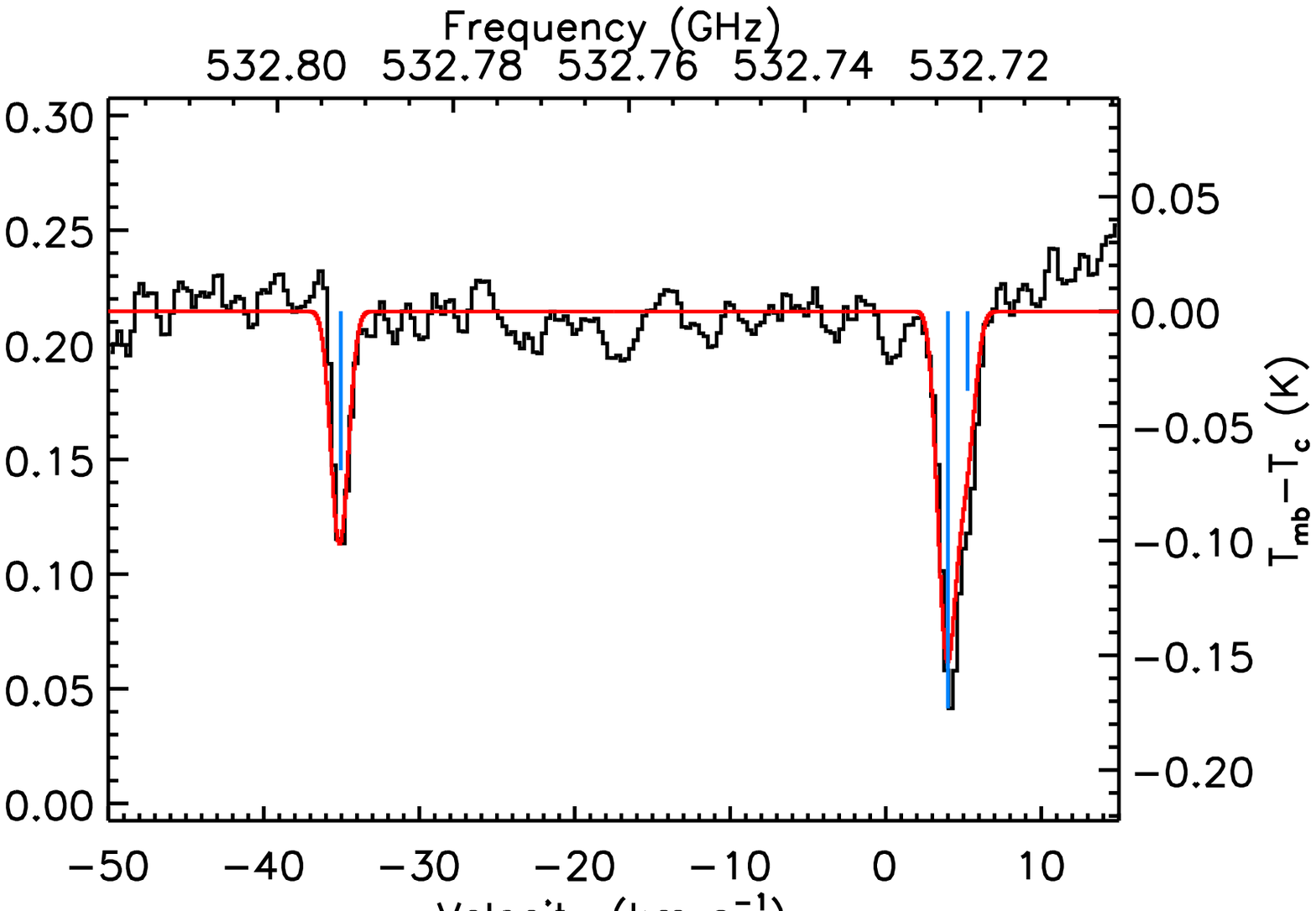}\hfill
\includegraphics[width=0.95\columnwidth]{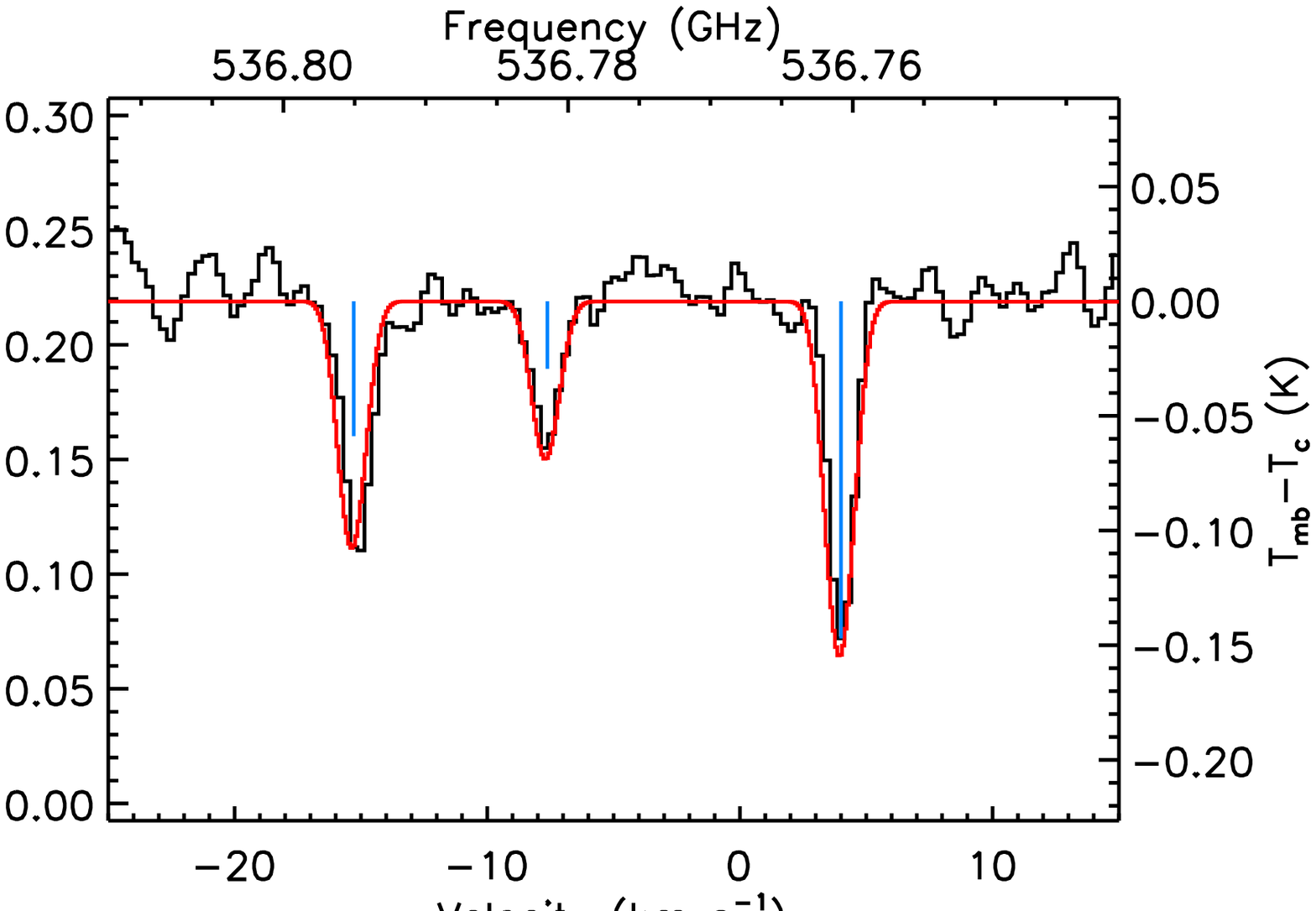}\\[2ex]
\includegraphics[width=0.95\columnwidth]{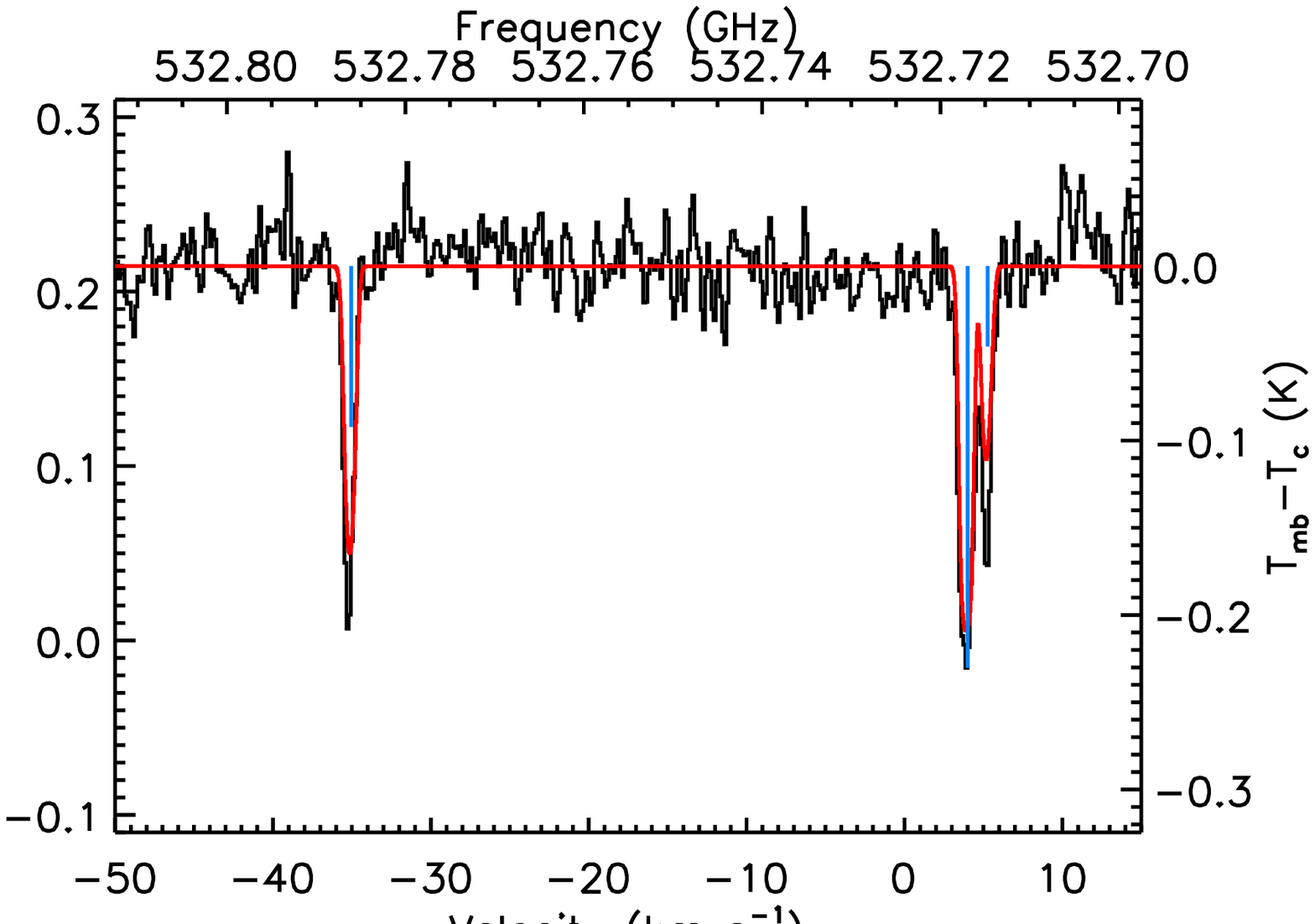}\hfill
\includegraphics[width=0.95\columnwidth]{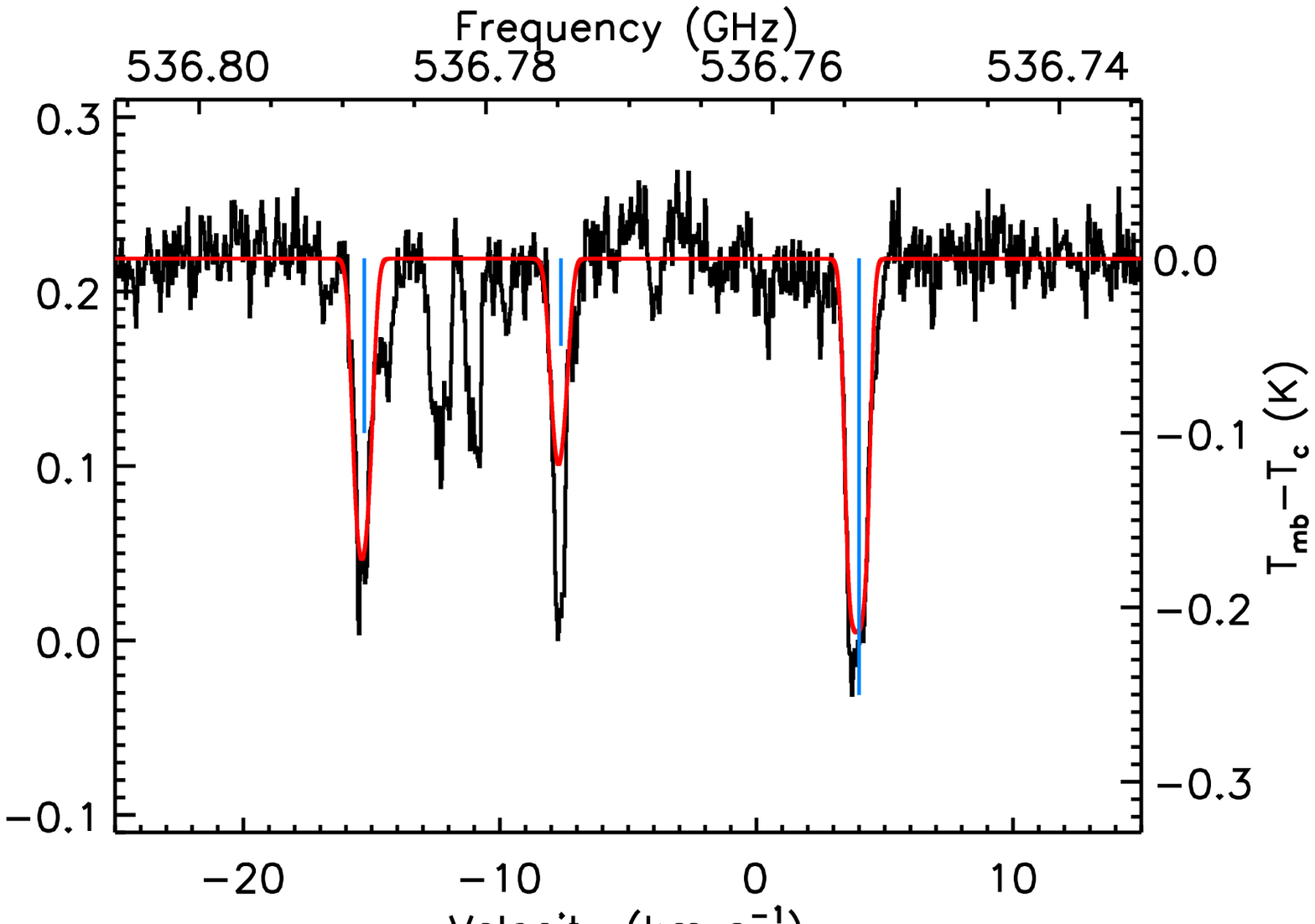}
\caption{CH ``survey'' spectra at $\sim$532 (top left) and $\sim536$ (top right) GHz, 
and CH ``pointed'' HRS spectra $\sim$532 (bottom left) and at $\sim536$~GHz (bottom right).
The spectra are overlaid with the result of the $\chi^2$ minimization (red) performed with CASSIS
on the HRS data ; for comparison with the survey data, the model was convolved to the appropriate spectral resolution of the WBS (1.1 MHz).
The blue lines show the expected relative line intensities in the optically thin case.
\label{fig: ch hfs fit} }
\end{figure*}

Figure~\ref{fig: ch hfs fit} shows the detection of the hyperfine structures (HFS) of
the $N = 1\leftarrow1$, $J = 3/2\leftarrow1/2$, $p=+ \leftarrow -$ and $p=- \leftarrow +$, transitions of CH, 
with their strongest components at the rest frequencies 532.72394 GHz and 536.7611~GHz respectively (see Table~\ref{tab: obs}).
Except for the pair at 532.72~GHz, the hyperfine components are resolved. 
Defining relative intensities as $\displaystyle r_i=\frac{A_{u\ell,i}g_{u,i}}{\sum_iA_{u\ell,i}g_{u,i}}$, where $g_{u,i}$ is the statistical weight of the upper level of transition $i$ and such that $\sum r_i =1$,
it can be seen from Fig.~\ref{fig: ch hfs fit} that the intensity ratios clearly deviate from what is expected (blue lines)
in the case of optically thin LTE excitation. More specifically, the lines with the highest $A_{u\ell}$ appear to be optically thick.\\
Figure~\ref{fig: ch hfs fit} also shows absorption features in the pointed HRS data that are not present in the survey WBS data ; 
moreover, the CH line at 536.7820\,GHz is deeper in the pointed than in the survey data.
We carefully checked (see Appendix~\ref{ap: contamination image band}) that these contaminations were due to contribution from the image band.\\

We derive CH line parameters 
by performing a $\chi^2$ minimization using the Monte Carlo Markov Chain (MCMC -- \citealt{guan-etal06,hastings70}) method in CASSIS\footnote{http://cassis.irap.omp.eu}
(Centre d'Analyse Scientifique de Spectres Instrumentaux et Synth\'etiques), a free-of-charge interactive spectrum analysis software package developed at IRAP.
The analysis presented here assumes LTE. Indeed, only approximate collisional coefficients are available in \citet{bouloy-etal84} ; moreover, new calculations for these rates are ongoing as part of the ANR\footnote{Agence Nationale pour la Recherche} ``HYDRIDES'' (contract ANR-12-BS05-0011-01, PI: A. Faure), which will combine a high-accuracy potential energy surface with coupled-channel calculations  (A. Faure, priv. comm.). We therefore postpone a non-LTE analysis to a future study.
The minimization explores parameters such as the column density ($N$), the excitation temperature (\tex), the full width at half maximum (FWHM), the source size and the local standard of rest velocity (\vlsr) of the fitted Gaussian (see Appendix \ref{ap: formalism}).
It was performed on the pointed/HRS data, 
and we only considered lines not contaminated by the image band.

Initial modeling of the CH lines indicated that the absorption occurred at $\sim$4~\kms, assuming that it originates from
a single physical component. We checked that the values obtained with CASSIS are consistent with those obtained
by fitting the hyperfine structure using the HFS method in the CLASS software 
(see the CLASS manual at http://www.iram.es/IRAMES/- otherDocuments/manuals/index.html and the description in
\citealt{bacmann-etal10} for more details).
However, this velocity corresponds neither to that of the envelope ($\sim$3.8~\kms), nor to that 
of a foreground cloud (see \S\ref{sec: intro}) responsible for the absorption at 4.2~\kms\ of a number of species
such as D$_2$O (\citealt{2013A&A...553A..75C,vastel-etal10}; see also Appendix~\ref{ap: ch vs d2o}),
HDO \citep{coutens-etal12},
ND (see Appendix~\ref{ap: ch vs d2o}).
Taking this into account, we performed a $\chi^2$ minimization assuming then that the absorption originates
from two physical components : the envelope and the foreground cloud revealed by the HDO observations of \citet{coutens-etal12}.
Given the small number of constraints,
we fixed as many parameters as possible.
From the $\chi^2$ minimization of the HRS observations of the D$_2$O and ND lines, we fixed 
the \vlsr\ of the foreground cloud to 4.2~\kms.
Additionally, only an upper limit of a few K could be found
for the excitation temperature of the two components, so we fixed both \tex\ at 2.73~K ; 
this assumption is consistent with the fact that collisional excitation is negligible because of the low gas density
($n\lesssim10^6-10^7$~\cc, the range of critical densities of the detected CH transitions) and moderate temperature.
Finally, at the low \tex\ that we assume, the emission is negligible, so that the extent of the envelope and of the foreground cloud, which are needed for CASSIS to perform the calculations, can be set to an arbitrary size.
Free parameters were then the column densities and FWHM of the two components, as well as the \vlsr\ of the envelope.

Best-fit parameters and 1-$\sigma$ uncertainties are reported in Table~\ref{tab: results} 
and the corresponding model spectra are overplotted on Fig.~\ref{fig: ch hfs fit}.
Note that we checked that the parameters in Table~\ref{tab: results} are consistent with the non-detections reported in Table~\ref{tab: obs},
assuming a [CD]/[CH] ratio of up to 1 (since deuterium fractionation ratio can be as large as 100\% in this source -- see, e.g., \citealt{bacmann-etal10})
and [$^{12}$CH]/[$^{13}$CH] = 69 \citep{wilson99}. 

The CH column density of $7\times10^{13}$~\cs\ observed in the envelope falls within the range of values
reported for dark clouds ($\sim2-15\times10^{13}$~\cs, e.g., \citealt{mattila86,jacq-etal87,sakai-etal07}), for which we expect similar physical parameters as for the envelope of \iras.

The lower CH column density derived for the foreground cloud, $1.5\times10^{13}$~\cs, is closer to the values reported for diffuse and translucent clouds\footnote{Translucent molecular clouds are defined by \citet{vandishoeck+black88} as interstellar clouds with $1 < {\rm A_V} < 5$.}
\citep[e.g.][]{vandishoeck+black86,liszt+lucas02,sheffer-etal08,chastain-etal10}, where
CH has been extensively observed.
These studies have found that, in the low-density regions where the chemistry is dominated by UV radiation,
such as diffuse clouds,
the CH column density correlates with the total molecular hydrogen column density ([CH]/[H$_2$]=$3.5\times10^{-8}$, \citealt{sheffer-etal08}). If diffuse cloud conditions pertained in our foreground cloud, converting N(CH) to N(H$_2$) to visual extinction would yield A$_{\rm V}\sim0.1$. 
However, we do not favor this.
Indeed, \citet{liszt+lucas02} noted that the relationship between N(CH) and extinction is bimodal: in diffuse molecular gas, N(CH) $\lesssim 3\times10^{12}$~\cs, while in translucent molecular gas, N(CH) $\gtrsim 10^{13}$~\cs, with the transition between low and high N(CH) occuring in the range $0.3\le {\rm A_V}\le1.2$. 
Considering the somewhat high galactic latitude of \iras\ ($b = 16^\circ$), it seems more likely that our foreground cloud would be such a translucent cloud rather than a diffuse cloud, more commonly found in lines of sight to the Galactic Center. This situation would also be consistent with \citet{coutens-etal12} who proposed the existence of an absorbing layer with A$_{\rm V} \sim 1-4$ to explain their HDO data.
Another point is that the study of \citet{chastain-etal10} on translucent clouds suggests that the CH column density does not always correlate with the H$_2$ column density, a conclusion also reached by \citet{sakai-etal07} for Heiles Cloud 2 in the Taurus molecular cloud complex. These authors interpret this as a consequence of the chemical evolutionary effects on the CH abundance.\\
Keeping this in mind, we investigate in the next section the chemistry of CH in the envelope of \iras\ and in the foreground cloud, assuming for the later that it is embedded in the $\rho$-Ophiucus molecular and has A$_{\rm V}\ge1$.

\begin{table}
\caption{Results of the $\chi^2$ minimization for the pointed/HRS data.\label{tab: results}}
\begin{center}
\begin{tabular}{cccc}
\hline\hline
Component$^a$ & $N$ & FWHM & \vlsr  \\ 
                               & ($10^{13}$~\cs) & (\kms)   & (\kms) \\
\hline
Envelope  & $7.7\pm0.2$ & $0.57\pm0.02$ & $3.87\pm0.01$ \\ 
Foreground cloud & $1.5\pm0.3$ & $0.46\pm0.02$ & 4.20$^b$ \\ 
\hline
\end{tabular}
\end{center}
$^a$ Excitation temperatures were set to 2.73 K for both the envelope and the foreground cloud. 
Both component are assumed to fill the beam.\\
$^b$ Fixed.
\end{table}

\section{Chemical modeling}
\label{sec: chem}

\subsection{Protostellar envelope}
\subsubsection{Model description}

To model the chemistry in \iras, we used the same approach as \citet{2008ApJ...674..984A}. In this model, the chemistry is computed with the Nautilus gas-grain code \citep{2009A&A...493L..49H,2010A&A...522A..42S}, which allows to compute the evolution of the chemical composition of the gas and the icy mantle of the grains. Compared to the previous applications with this code, the gas-phase and grain surface networks have been changed. The gas-phase is now based on the kida.uva.2011 network \citep[see][]{2012ApJS..199...21W}, which corresponds to the merging of the OSU network (http://www.physics.ohio-state.edu/$\sim$eric/research.html) into the KIDA database (http://kida.obs.u-bordeaux1.fr/). In addition, a few gas-phase reactions have been added for species specific to the grain surface chemistry (for O$_3$, C$_n$H$_3$ and C$_n$H$_4$ for instance). The surface network is based on the network developed by Prof. Eric Herbst's team (University of Virginia, USA) \citep{2007A&A...467.1103G}. Some very minor species have been removed however in order to simplify the network without changing significantly the model predictions (mostly H$_2$C$_n$N and H$_3$C$_n$N species). In the end, our full network contains 684 species (485 in the gas-phase and 199 at the surface of the grains) and 7957 reactions (6177 pure gas-phase reactions and 1780 reactions of interactions with grains and reactions at the surface of the grains). The full network is available on the KIDA webpage: http://kida.obs.u-bordeaux1.fr/models. \\

The physical structure of the protostellar envelope has been computed with the 1D radiation hydrodynamic model (RHD) of a spherical collapsing core by \citet{2000ApJ...531..350M}. In this model, the initial H$_2$ density is $3\times 10^4$~cm$^{-3}$ and the total mass is 3.852~M$_\odot$. After an isothermal contraction, the compressional heating overwhelms the cooling, producing an increase of the temperature towards the center of the envelope. In this model, it takes $2.5\times 10^5$~yr to evolve from a pre-stellar to a protostellar core. After the second collapse, which represents the birth of the protostar, the model follows the evolution of the physical conditions for $9.3\times 10^4$~yr. In practice, the envelope is divided in 14 different shells. For each of them, we have the radius, the temperature, the density and the visual extinction at each time step. These pieces of information are used as input parameters of our chemical model. There is no feedback of the chemical calculations on the physical structure. The final temperature and density profiles in our simulations are very similar to the ones from \citet{2010A&A...519A..65C}, derived from an analysis of multi-wavelength continuum observations towards \iras\ (see Fig. 1 and Table 1 of \citealt{2010A&A...519A..65C}). Only the density profile is approximately ten times smaller (at all radii). To be more consistent with the densities used for the analysis of the observations, we have multiplied all densities (of the 14 shells) by a factor of ten for the calculation of the chemistry as a function of time. In the section presenting the results, we discuss the importance of this assumption. The dust temperature is assumed to be equal to the gas temperature. More details on the physical model can be found in \citet{2008ApJ...674..984A}. \\

In addition to this period of collapse, which is done in two steps ($2.5\times 10^5$~yr for the evolution of the pre-stellar to the protostellar core followed by the growth of the protostellar envelope during $9.3\times 10^4$~yr), we assume that the pre-stellar core remains static during $10^6$~yr before starting to collapse. During this period, each shell has the physical conditions stated by the physical model, i.e. a temperature of about 7~K and a density of about $2\times10^4$~\cc, so that the initial composition is homogeneous. In total, the chemistry is calculated during $1.343\times 10^6$~yr. We used the same elemental abundances as \citet[][Table 1, with the oxygen elemental abundance of $3.3\times10^{-4}$ compared to H, which gives an elemental C/O ratio of 0.5]{2011A&A...530A..61H}. All species are assumed to be initially in the atomic form except for hydrogen, which is initially already molecular. The photodissociations are treated in 1D, assuming only the external interstellar UV field of 1~G$_0$. The visual extinction in the envelope is then a function of the hydrogen column density (A$_{\rm V}$ = N$_{\rm H}$/(1.59$\times$10$^{21}$\cs)) and we add 3 to this computed A$_{\rm V}$ to take into account the additional extinction from the giant molecular cloud in which we assume that the object is embedded \citep{2008ApJ...674..984A}. We compute the CO and H$_2$ photodissociation rates as a function of H$_2$ and CO column densities and as a function of the visual extinction using the approximation from \citet{1996A&A...311..690L}. Based on \citet{doty-etal04}, we have used a cosmic-ray ionization rate, $\zeta$, of 10$^{-16}$~s$^{-1}$ but we also run the model for $\zeta = 10^{-17}$~s$^{-1}$.

\subsubsection{Model result}\label{proto_results}

\begin{figure}
\includegraphics[width=0.95\columnwidth]{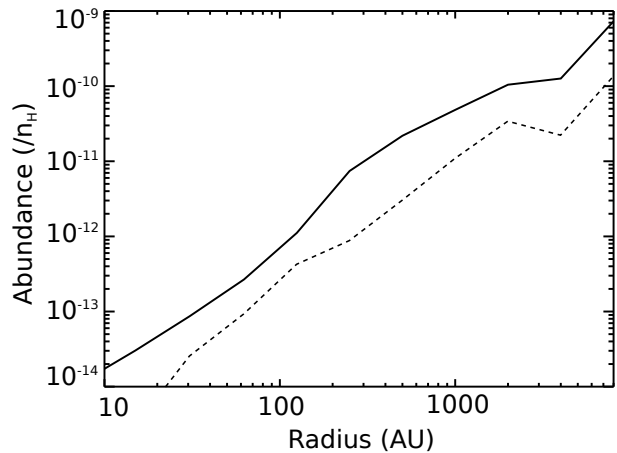}
\caption{CH abundance profile in the protostellar envelope as a function of the radius to the center predicted by our model for a cosmic-ray ionization rate of $10^{-16}$ (solid line) and  $10^{-17}$~s$^{-1}$ (dashed line), $9.3\times10^4$~yr after the birth of the protostar (see text for details). 
\label{CH_dynamic} }
\end{figure}

Figure \ref{CH_dynamic} shows the CH abundance profile predicted by our model at the end of the simulations for the parameters described in the previous section (solid line). The CH abundance increases towards the outer radii up to an abundance of $7\times 10^{-10}$ at 8000 AU (compared to the total proton density). The external abundance is set by the pre-collapse phase whereas the decrease of the CH abundance towards the center is a consequence of the increase of the temperature. The resulting CH column density is $2\times 10^{13}$~cm$^{-2}$, i.e. about 3.8 times smaller than the observed one. Here, we only consider half of the envelope since the signal is observed in absorption. \\
Using a smaller cosmic-ray ionization rate of $1.3\times 10^{-17}$~s$^{-1}$ (closer to the one most commonly used for dense environments) produces less CH in the entire envelope as can be seen from Fig. \ref{CH_dynamic} (dashed curve). The CH column density in this case in the envelope is $5\times 10^{12}$~cm$^{-2}$.\\
Using the original densities computed by the RHD model without the factor of 10 increase, the CH abundances predicted by the model are slightly larger but the column densities are approximately the same in the case of $\zeta = 10^{-17}$ and two times smaller for the higher $\zeta$. Whatever the cosmic-ray ionization rate, if we consider that the pre-stellar core stays static over $10^5$~yr, instead of $10^6$~yr, our model predictions are not significantly changed. 

\subsection{Foreground cloud}

\begin{figure}
\includegraphics[width=0.95\columnwidth]{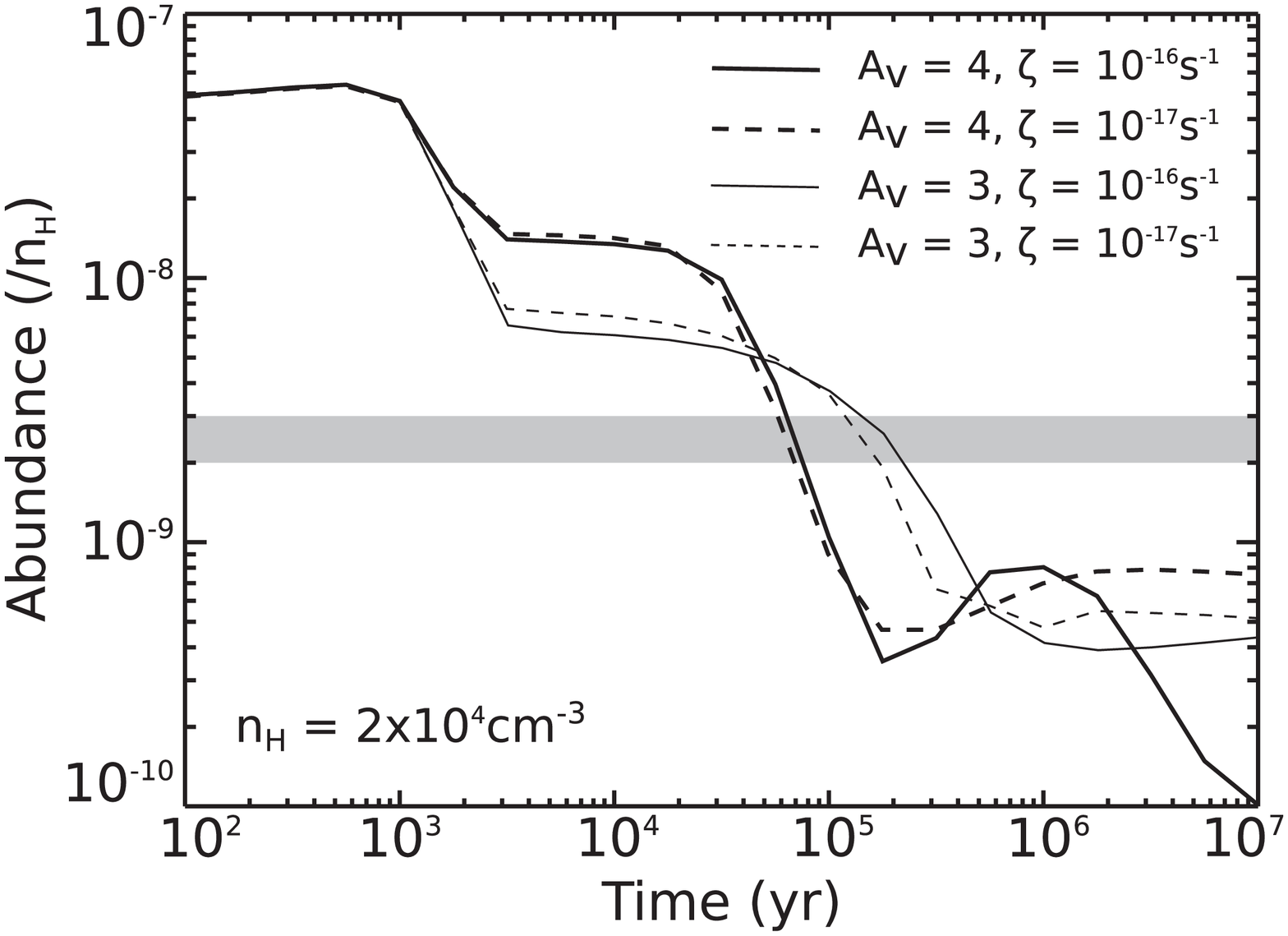}
\includegraphics[width=0.95\columnwidth]{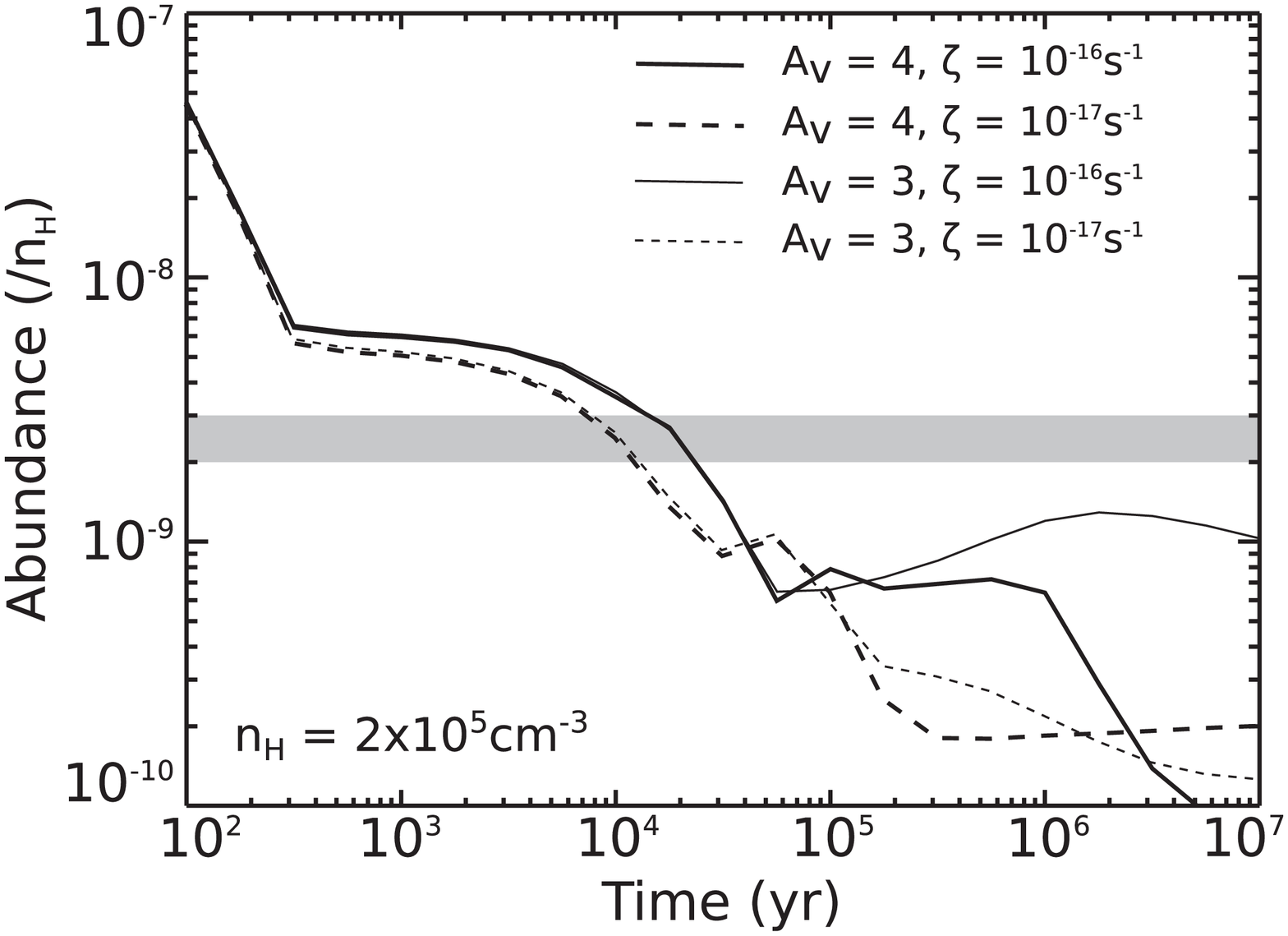}
\caption{CH abundance as a function of the time for the foreground cloud predicted by our model. 
Solid and dashed lines have been obtained with a cosmic-ray ionization rate of $10^{-16}$ and $10^{-17}$~s$^{-1}$ respectively whereas thin and thick lines indicate an A$_{\rm V}$ of 3 and 4 respectively. 
The upper panel is for a H density of $2\times 10^{4}$~cm$^{-3}$ whereas the lower panel is for a H density of $2\times 10^{5}$~cm$^{-3}$.
The range of CH abundances corresponding to the observed column density in the foreground cloud of $1.5\times10^{13}$~\cs\ is shown in grey.
\label{CH_cloud} }
\end{figure}

To model the CH abundance in the foreground molecular cloud, we have used Nautilus in 0D for fixed physical conditions and times up to $10^7$~yr. 
The initial conditions are the same as for the pre-stellar static core described in the previous section: atomic composition with abundances from \citet{2011A&A...530A..61H} except for hydrogen, which is initially molecular. Note that the timescales for the chemical evolution of this cloud are distinct from the ones for the protostellar envelope if the two sources are spatially separated. If not, time = 0 for the foreground cloud should correspond to time = 0 for the static pre-stellar phase of the previous modeling. \\
Considering the weak constraints on the physical conditions of this source, we have run a large grid of chemical models and used the observational constraints on the singly and doubly deuterated water from \citet{2013A&A...553A..75C}, which was the first observational study to highlight this foreground layer. Temperatures of 15 and 30~K, total proton densities of $2\times 10^4$ and $2\times 10^5$~cm$^{-3}$, A$_{\rm V}$ between 1 and 4 and cosmic-ray ionization rate of $10^{-17}$ and $10^{-16}$~s$^{-1}$ have been considered. \\
Around $10^5$~yr, the CH molecule is mostly produced by the dissociation of CH$_2$, CH$_3$ and CH$_4$ by direct UV photons and photons induced by cosmic-rays, and the reaction H + CH$_2$ $\rightarrow$ CH + H$_2$ (whose rate coefficient is temperature-independent). The main destruction mechanisms are the direct UV and cosmic-ray induced UV photons as well as the reaction H + CH $\rightarrow$ C + H$_2$. Smaller A$_{\rm V}$ or larger $\zeta$ directly increase the photodissociation processes whereas larger temperatures produce larger abundances of atomic hydrogen. A temperature above 15~K clearly decreases the CH abundance.\\
 In Fig.~\ref{CH_cloud}, we show the CH abundance computed by the model as a function of time for a temperature of 15~K only but for two different densities ($2\times 10^4$ and $2\times 10^5$~cm$^{-3}$), two different A$_{\rm V}$ (3 and 4) and two different cosmic-ray ionization rates ($10^{-17}$ and $10^{-16}$~s$^{-1}$). 
\citet{coutens-etal12} proposed that this foreground layer should have an A$_{\rm V}$ between 1 and 4. However, at A$_{\rm V} \lesssim 3$, photodissociation dominates and the predicted CH abundances are negligible, and we therefore exclude these low values of A$_{\rm V}$.
Using the H column densities corresponding to A$_{\rm V}$ of 3 -- 4 (N$_{\rm H} \sim 5\times10^{21}$~\cs),
we obtain a range of predicted CH column densities between $5\times10^{11}$ and $2.5\times10^{14}$~\cs,
which largely encompasses the observed value of $1.5\times10^{13}$~\cs.
For an easier comparison, we consider, in the analysis that follows,
the observed abundance of CH derived from the observed CH column density and the above value for N$_{\rm H}$: we obtain approximatively $(2-3)\times10^{-9}$,
shown in grey in Fig.~\ref{CH_cloud}. \\ 
%
%
For a density of $2\times10^4$~\cc\ and a given A$_{\rm V}$, we get similar results whatever the cosmic-ray ionization rate for times smaller than a few $10^5$~yr. For A$_{\rm V}$ = 4, the model reproduces the observations for times of about $7\times10^4$~yr. For A$_{\rm V}$ = 3, the agreement time is slightly larger: $(1-2)\times10^5$~yr. At higher density, the agreement is obtained at earlier times, at which the model predictions do not depend on the visual extinction. The observations are reproduced between $7\times10^3$ and $10^4$~yr if $\zeta = 10^{-17}$~s$^{-1}$ and between $10^4$  and $2\times10^4$~yr for $\zeta = 10^{-16}$~s$^{-1}$. 


\subsection{Discussion}

\subsubsection{Timescales}

Ages of protostars or molecular clouds are very often claimed in papers based on the dynamics of the regions or the chemical composition. For the protostellar envelope of \iras, many estimates are available that range from a few $10^3$~yr to a few $10^4$~yr. Indeed, for protostellar envelopes, two types of ages can be found. The first one is based on the study of the dynamics of the region through the adjustment of the parameters of an inside-out collapse model to fit line profiles. Using such method, \citet{stark-etal04} obtained a dynamical age of $(0.6-2.5)\times10^4$~yr and \citet{schoier-etal02} obtained $(1-3)\times10^4$~yr for \iras. The time = 0 for this age should be the start of the collapse. Compared to our approach, it should correspond to the end of the static stage. In our case, the free-fall timescale is much longer due to the fact that the initial density is smaller than the density observed in the outer parts of the protostellar envelope. Another way to estimate the age of the protostar is to compare the observed chemical composition in the inner hot region of the envelope with time-dependent chemical models. The age obtained in this case correspond to a ``chemical age'' with time = 0 the time at which the grain mantles have evaporated in the gas-phase. Such estimate is quite model dependent. \citet{cazaux-etal03} for instance found an age of $5\times10^4$~yr by comparing NH$_3$/CH$_3$OH abundance ratios with model predictions by \citet{rodgers+charnley01}. \citet{wakelam-etal04} found a much smaller chemical age of $2\times10^3$~yr by studying the sulphur chemistry.\\
Since the CH molecule is located in the outer parts of the protostellar envelope where the physical conditions do not evolve much, we showed that its abundance does not depend much on the timescales and the density. 
Concerning the foreground cloud, if it is dynamically connected to the envelope of \iras\ then the early age corresponding to the models with n$_{\rm H} = 2\times10^5$~\cc\ (which is the density in the outer part of the protostellar envelope) is consistent with previous estimates of the age of the protostar. However, if we interpret the different \vlsr\ of the envelope and the foreground cloud as an indication that the two are not dynamically connected, then it is conceivable that the foreground cloud formed earlier (and has a lower density compared to the protostellar envelope), at the same time as the $\rho$-Ophiucus giant molecular cloud whose estimated age ranges between 0.1 and 1 Myr, with a median age of $3\times10^5$~yr \citep{greene+meyer95,luhman+rieke99}. 
  
\subsubsection{Importance of the reaction H + CH $\rightarrow$ C + H$_2$}

From the chemical modeling that has been presented in this paper, we found that the CH abundance was particularly sensitive to the adopted rate coefficient of the CH destruction reaction H + CH $\rightarrow$ C + H$_2$. The rate coefficient that we have used has the temperature following temperature dependence: $\rm k(T) = 1.24\times 10^{-10} (T/300)^{0.26}$~ cm$^3$~s$^{-1}$, which gives a rate coefficient of $5\times 10^{-11}$~cm$^3$~s$^{-1}$ at 10~K. This value is the one recommended by the KIDA database\footnote{http://kida.obs.u-bordeaux1.fr}. Compared to the values previously used, it has been increased by a factor of 7, reducing the predicted abundance of CH. Using the previous lower value of the rate coefficient, the predicted abundance of CH in the outer part of the envelope of the protostar (see section~\ref{proto_results}) is multiplied by more than a factor of 5 and the total CH column density is $7\times 10^{13}$~cm$^{-2}$. 
The effect on the chemistry of the foreground cloud is also strong. As an example, using a density of 2$\times10^4$~\cc, an A$_{\rm V}$ of 3 and a cosmic-ray ionization rate of 1$\times10^{-16}$~s$^{-1}$, the predicted CH abundance would always be larger than 2$\times10^{-9}$ (compared to H). \\
The KIDA recommendation is based on a few experimental measurements and theoretical calculations but is quite uncertain according to the KIDA experts. Further investigation on the low temperature estimate of this rate coefficient may be useful.
The importance of this reaction depends on the abundance of atomic hydrogen in our simulations.  The H abundance in the gas-phase is about $10^{-5}$ (compared to the total proton density) in the entire protostellar envelope if $\zeta = 10^{-17}$~s$^{-1}$ whereas it is approximately ten times larger for $\zeta = 10^{-16}$~s$^{-1}$.  In the foreground cloud with a density of $2\times10^4$~\cc, all models predict that the abundance of hydrogen in the gas-phase increases with time with an abundance of approximately $10^{-4}$ except for the model with the $\zeta = 10^{-16}$~s$^{-1}$ and A$_{\rm V}$ = 4 which produces a seven times larger H abundance at that time. At higher density, the gas-phase H abundance shows a decrease with time (except the model with A$_{\rm V}$ = 4 and $\zeta = 10^{-16}$~s$^{-1}$ where H increases strongly with time).  The predicted abundance is between $4\times10^{-5}$ and $10^{-4}$ at $10^4$~yr for all models. Atomic hydrogen is mainly produced by the direct dissociation of H$_2$ by cosmic-rays. \\

\section{Conclusions}

We presented {\it Herschel}/HIFI observations of the CH ground-state transitions at $\sim$532 and $\sim$536~GHz towards the protostar IRAS162932, obtained as part of the CHESS guaranteed time key program, which we complemented with archive data. Using the MCMC $\chi^2$ minimization in CASSIS, the line profiles can be reproduced assuming LTE and that CH originates both from the protostellar envelope, which has a \vlsr\ of $\sim$3.9~\kms, and from a foreground cloud with a \vlsr\ of 4.2~\kms, consistent with the results obtained for other species seen in absorption in this source \citep[e.g., ][]{vastel-etal10,coutens-etal12}. 
This minimization yielded CH column 
densities of $(7.7\pm0.2)\times10^{13}$~\cs\ and $(1.5\pm0.3)\times10^{13}$~\cs\ in the envelope and foreground cloud respectively.\\
Chemical modeling was performed using the Nautilus gas-grain code and the kida.uva.2011 network. 
Our model predicts that the CH abundance peaks at the outer layer of the protostellar envelope in agreement with the observations. The modeled column density is 3.8 times smaller than the observed one using a large ionization cosmic-ray rate of $10^{-16}$~s$^{-1}$ as suggested by \citet{schoier-etal02}. A smaller $\zeta$ worsens the agreement. A study of the uncertainty propagation of the model parameters (see \citealt{wakelam-etal06}) would be needed to define error bars on the model predictions and conclude on the agreement with the observations. Such studies have however never been done up to now for gas-grain models. In the foreground cloud, the observations are reproduced by our model at early times ($\sim 10^4$~yr) if the density is high ($2\times 10^5$~cm$^{-3}$). In that case, the model results show little sensitivity to the visual extinction whereas it is sensitive to the cosmic-ray ionization rate. The larger $\zeta$, the larger the time of agreement. In the case of a less dense cloud ($2\times 10^4$~cm$^{-3}$), the observations are reproduced by the model at later times ($\sim 10^5$~yr). In that case, the model results show little dependence on $\zeta$ whereas smaller A$_{\rm V}$ produces larger times of agreement. 
\\
In all our models, the CH abundance is very sensitive to the reaction of destruction H + CH $\rightarrow$ C + H$_2$.
As KIDA experts deem its rate coefficient to be uncertain, further investigation on this parameter at low temperature is necessary in order to obtain a more accurate modeling of the envelope of IRAS16293. Additionally, upcoming, more accurate collisional rate coefficients will allow us to perform non-LTE calculations, and hence to derive more stringent constraints for the chemical model.

\section*{Acknowledgments}

HIFI has been designed and built by a consortium of institutes and university departments from across Europe, Canada and the United States under the leadership of SRON Netherlands Institute for Space Research, Groningen, The Netherlands and with major contributions from Germany, France and the US. Consortium members are: Canada: CSA, U.Waterloo; France: IRAP (formerly CESR), LAB, LERMA, IRAM; Germany: KOSMA, MPIfR, MPS; Ireland, NUI Maynooth; Italy: ASI, IFSI-INAF, Osservatorio Astrofisico di Arcetri-INAF; Netherlands: SRON, TUD; Poland: CAMK, CBK; Spain: Observatorio Astron\'omico Nacional (IGN), Centro de Astrobiolog\'{i}a (CSIC-INTA). Sweden: Chalmers University of Technology - MC2, RSS \& GARD; Onsala Space Observatory; Swedish National Space Board, Stockholm University - Stockholm Observatory; Switzerland: ETH Zurich, FHNW; USA: Caltech, JPL, NHSC. We thank many funding agencies for financial support.\\
VW's research is funded by the ERC Starting Grant 3DICE (grant agreement 336474).\\
Pointed observations of CH at $\sim$532~GHz (ObsId 1342227403 and 1342227404) were part of the Open Time 1 program led by Pierre Hily-Blant.

\bibliographystyle{mn2e} 
\bibliography{bib_sb,bib_val}

\appendix

\section[]{Contamination from the image band in the pointed observations}
\label{ap: contamination image band}

\begin{figure*}
\includegraphics[width=0.475\textwidth]{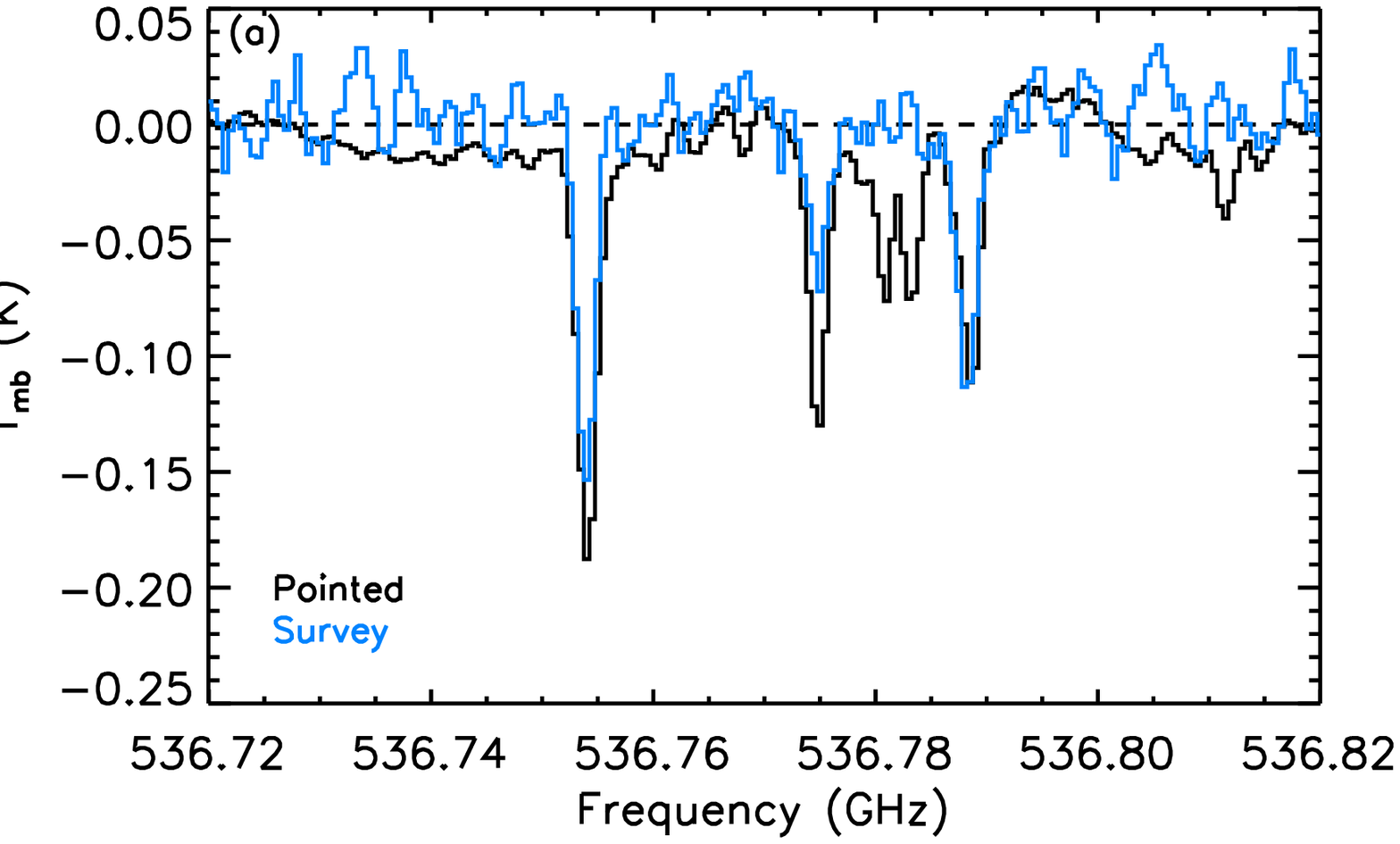}\hfill%
\includegraphics[width=0.475\textwidth]{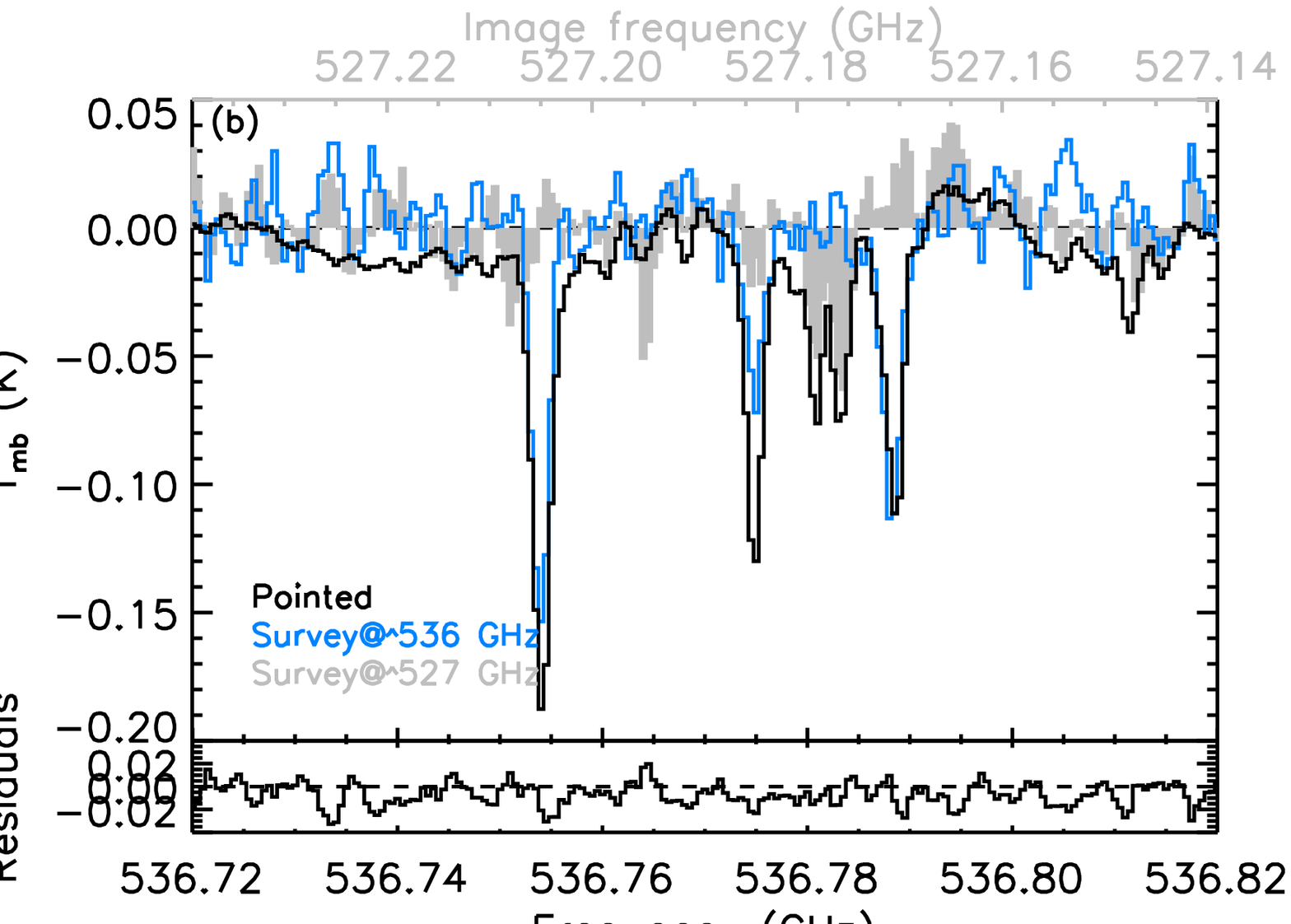}
\caption{(a) Comparison of the WBS spectrum obtained in the pointed observation of CH at $\sim$536~GHz (black)
and of the result of the deconvolution of the survey data over the same frequency range (blue).
--- (b) Same as (a), but with the additional overlay of the result of the deconvolution of the survey data over the frequency range corresponding
to the image band of the pointed observation (filled light grey). The residuals obtained by subtracting the sum of the blue and grey spectra from the black spectrum
are displayed in the bottom panel.
\label{fig: survey vs pointed}}
\end{figure*}

The comparison between the survey WBS and pointed HRS spectra (see right panels of Fig.~\ref{fig: ch hfs fit})
reveals the presence, in the pointed data, 
of contaminating absorption lines coming from the image band, 
resulting in a deeper than expected middle component of the CH triplet.
This contamination is also seen in the pointed WBS data, as illustrated in Fig.~\ref{fig: survey vs pointed}-a,
which shows the final pointed WBS spectrum (black) overlaid with the result of the deconvolution (blue).

To verify that this deepening is solely due to contamination from the image band, we added 
the survey spectrum taken at frequencies corresponding to the signal band of the pointed observations (blue spectrum in Fig.~\ref{fig: survey vs pointed}-b) to 
the survey spectrum taken at frequencies corresponding to the image band of the pointed observations (green spectrum in Fig.~\ref{fig: survey vs pointed}-b).
The resulting spectrum was subtracted from the pointed WBS spectrum (black spectrum in Fig.~\ref{fig: survey vs pointed}-b), and the residuals we obtained
are shown in the bottom panel of Fig.~\ref{fig: survey vs pointed}-b. These residuals are consistent with noise, which
indicates that the extra features in the pointed spectrum are indeed due to absorptions present in the image band. 
Note that these absorption lines remain unidentified since no transitions
at these frequencies (527.1837, 527.1863, and 527.1919~GHz), could be found in available catalogs (CDMS, JPL).

\section[]{Formalism used in CASSIS}
\label{ap: formalism}

In this section, we briefly explain how the synthetic LTE spectra are calculated in CASSIS.
The equations presented here are adapted from the document written by C. Vastel describing the LTE model\footnote{http://cassis.irap.omp.eu/docs/RadiativeTransfer.pdf} ; for more details, we refer the user to this document that can be found in
the documentation section of the CASSIS website\footnote{http://cassis.irap.omp.eu/?page=documentation}.\\

Let's consider a single species. For a Gaussian line shape, the opacity of this species as a function of velocity is given by:
\begin{equation}
\tau(\varv)=\sum_i \tau_{0,i}\exp\left(-\frac{(\varv-\vlsr)^2}{2\sigma^2}\right),\label{eq: tau}
\end{equation}
where:
\begin{myitemize}
\item $\tau_{0,i}$ is the opacity of transition $i$ at the line center and is given by:\\
\begin{equation}
\tau_{0,i}=\frac{1}{4\pi}~\sqrt{\frac{\ln2}{\pi}}~\frac{c^3A_{u\ell,i}g_{u,i} N}{\nu_{0,i}^3\Delta\varv~Q(\tex)\e^{E_{u,i}/k\tex}}(\e^{h\nu_{0,i}/k\tex}-1),
\end{equation}
with 
$A_{u\ell,i}$ the Einstein A-coefficient for spontaneous emission of transition $i$, 
$g_{u,i}$ the degeneracy of the upper level of transition $i$, 
$N$ the column density of the species of interest, 
$\nu_{0,i}$ the frequency of transition $i$, 
$\Delta\varv$ is the FWHM in velocity units, 
\tex\ the excitation temperature, 
$Q(\tex)$ the partition function at \tex, 
$E_{u,i}$ the energy of the upper level of transition $i$.

\item \vlsr\ is the source's velocity in the local standard of rest.

\item $\displaystyle \sigma(\kms) = \frac{\Delta\varv(\kms)}{2\sqrt{2\ln2}}$.\\
\end{myitemize}

For a single component, assuming no contribution from dust,
CASSIS calculates the LTE spectrum (brightness temperature) with the following equation:
\begin{equation}
T_{b,1} = T_C\e^{-\tau_1} + [ \Omega_1 J_\nu({\tex}_{,1})-J_\nu(\tcmb)](1-\e^{-\tau_1})
\end{equation}
where:
\begin{itemize}
\setlength{\leftmargin}{2.75cm}
\item $T_C$ is the temperature of the continuum.
\item $\tau_1$ is the opacity given by (\ref{eq: tau}).
\item $\Omega_1$ is the dilution factor given by $\frac{\theta_1^2}{\theta_1^2+\theta_b^2}$, where $\theta_1$ is the spatial extent of the component, and $\theta_b$ the half-power beam width of the telescope, both in arcseconds.
\item $\displaystyle J_\nu(T) = \frac{h\nu/k}{\e^{h\nu/kT}-1}$, $h$ and $k$ being Planck's and Boltzmann's constant, respectively.
\end{itemize}

Finally, for two components, assuming that component 2 is in front of component 1, CASSIS calculates the LTE spectrum with the following equation:
\begin{equation}
T_{b,2} = T_{b,1}\e^{-\tau_2} + [ \Omega_2 J_\nu({\tex}_{,2})-J_\nu(\tcmb)](1-\e^{-\tau_2})
\end{equation}

\section[]{Comparison with other absorption lines}
\label{ap: ch vs d2o}

\begin{figure}
\includegraphics[width=\columnwidth]{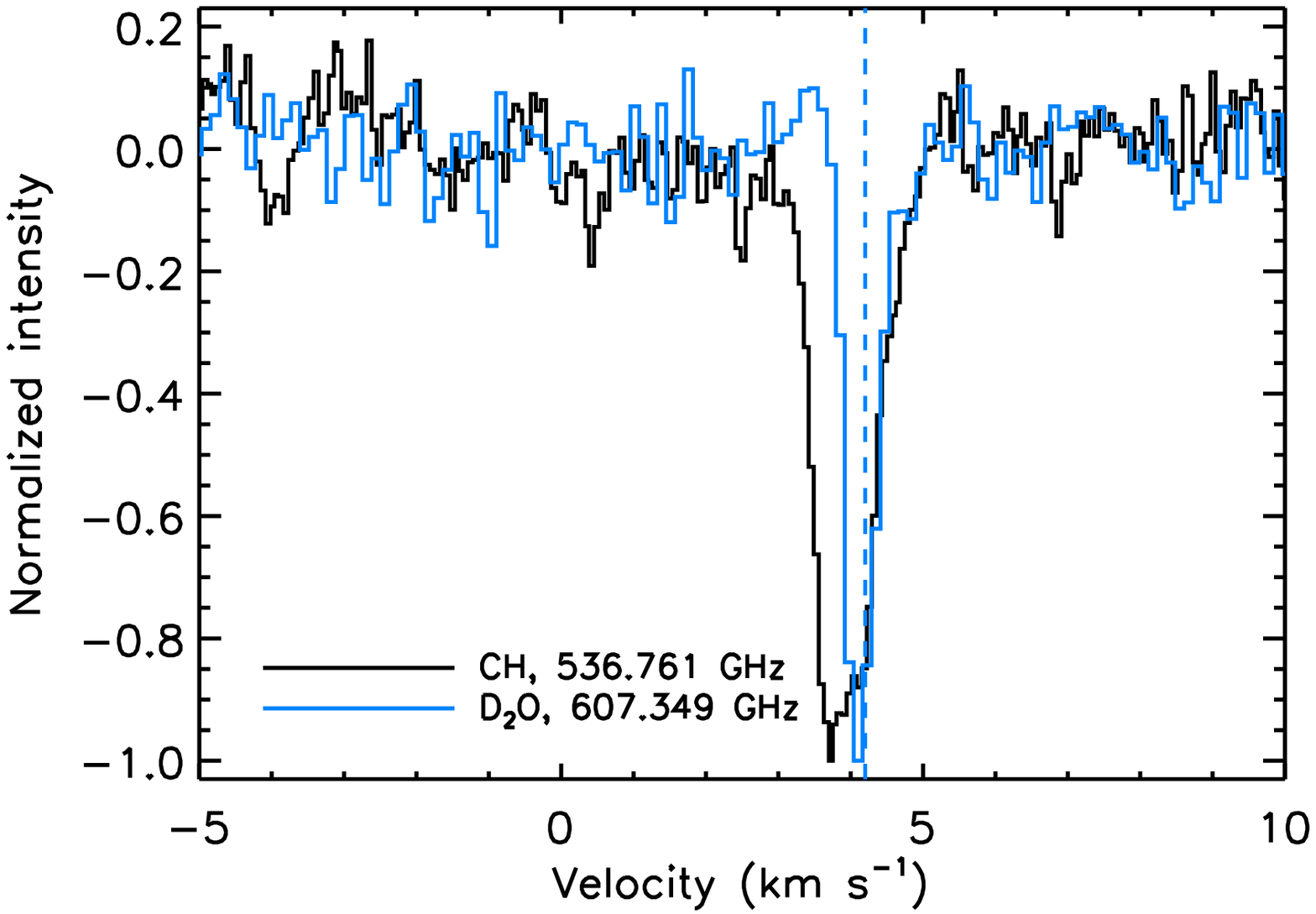}
\includegraphics[width=\columnwidth]{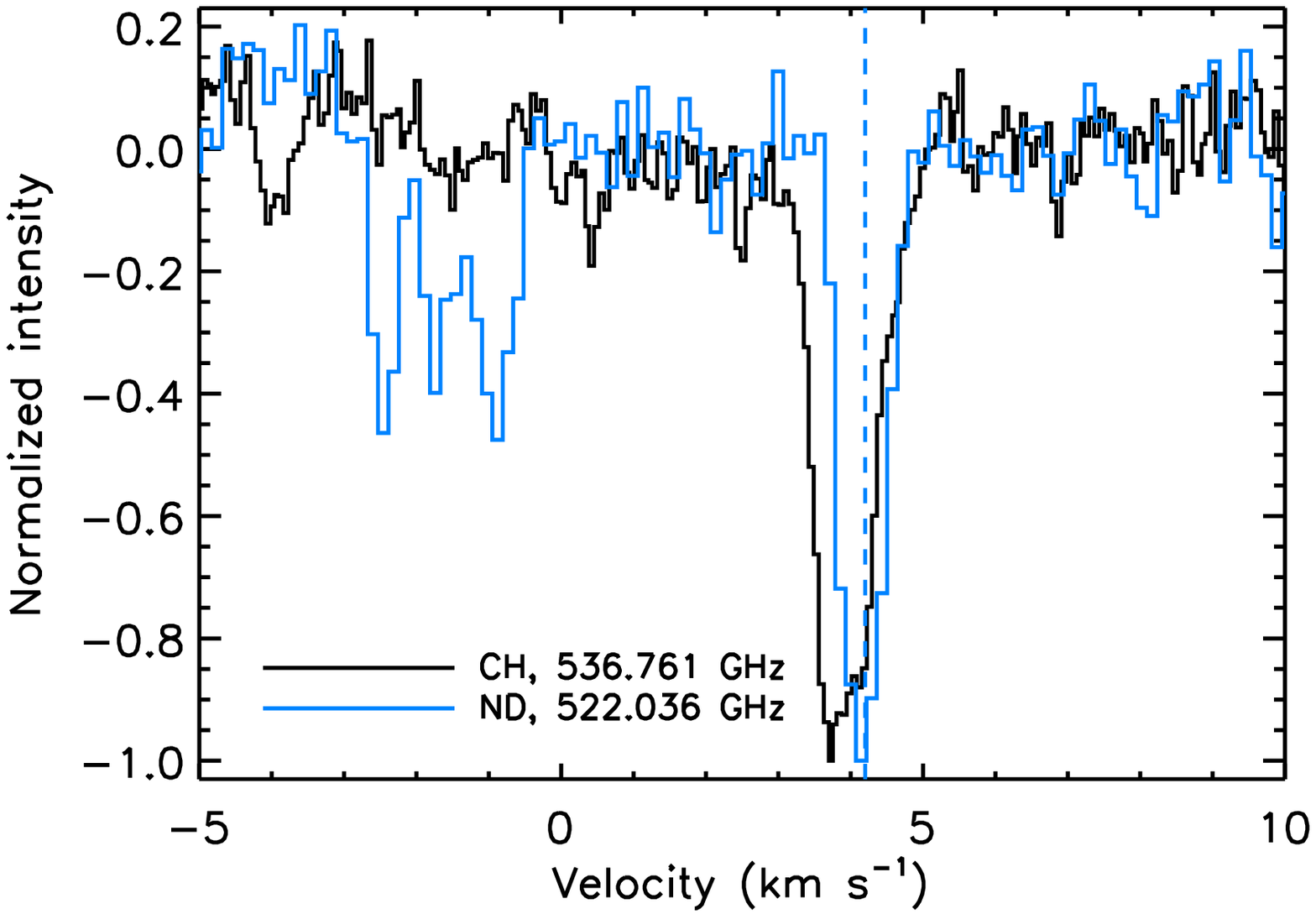}
\caption{CH transition at 536.761 GHz (black) overlaid with D$_2$O at 607.349 GHz (blue, top panel)
and with ND at 522.036 GHz (blue, bottom panel).
Each spectrum is continuum-subtracted and then normalized to the maximum depth of the absorption.
The vertical dashed red line indicates the position, 4.2~\kms, obtained with the $\chi^2$ minimization
of the red spectra in CASSIS.
\label{fig: ch vs d2o}}
\end{figure}

Initial modeling of the CH lines indicated that the absorption should lie at $\sim$4~\kms, assuming that it originates from
a single physical component. 
This velocity corresponds neither to that of the envelope ($\sim$3.8~\kms), nor to that 
of a foreground cloud at 4.2~\kms\ revealed by \citet{coutens-etal12}. 
This cloud is responsible for the self-absorption seen in some of the HDO lines studied by these authors, 
and for the absorption observed in other species such as D$_2$O \citep{vastel-etal10} or ND.
Indeed, using CHESS pointed observations of D$_2$O at 607.349~GHz and of the hyperfine structure of ND at $\sim$522~GHz, we performed a $\chi^2$ minimization on the HRS data of these lines. The results, a \vlsr\ of 4.2~\kms, and narrow widths of 0.4-0.45~\kms, are characteristic of the foreground cloud.
To illustrate the difference in line profiles, and support our choice of modeling CH with two physical components (the envelope and the foreground cloud),
we show in Fig.~\ref{fig: ch vs d2o} a comparison of the CH absorption line at 536.761~GHz with the CHESS pointed observations of D$_2$O at 607.349~GHz and ND at 522.036~GHz.  
The CH line profile encompasses the D$_2$O and ND profiles, consistent with our assumption that CH originates both from the envelope and the foreground cloud.

\bsp

\label{lastpage}

\end{document}